\documentclass[12pt]{article}
\usepackage{amsmath,graphicx}
\usepackage{bbm}
\usepackage{amssymb}
\usepackage{cite}                   
\usepackage{epsf}
\usepackage{epsfig}
\usepackage{psfrag}
\usepackage{rotating}                              
\usepackage{color}       
\newcommand{\lsim}{
\mathrel{\hbox{\rlap{\hbox{\lower4pt\hbox{$\sim$}}}\hbox{$<$}}}}

\newcommand{\gsim}{
\mathrel{\hbox{\rlap{\hbox{\lower4pt\hbox{$\sim$}}}\hbox{$>$}}}}

\newcommand{\ev}{\, {\rm eV}}

\newcommand{\bea}{\begin{eqnarray}}
\newcommand{\eea}{\end{eqnarray}}
\newcommand{\be}{\begin{equation}}
\newcommand{\ee}{\end{equation}}
\newcommand{\bi}{\begin{itemize}}
\newcommand{\ei}{\end{itemize}}

\renewcommand{\baselinestretch}{1.2}

\begin{document}
%
%
%
%
%
%
%
\vspace*{-0.5truecm}

\begin{flushright}
TUM-HEP-655/06\\
\end{flushright}


\begin{center}
\boldmath

{\Large{\bf Minimal Lepton Flavour Violation and Leptogenesis with
    exclusively low-energy CP Violation}}
\unboldmath
\end{center}


\begin{center}
\small

{\bf Selma Uhlig}

\vspace{0.5truecm}

{\sl Physik Department, Technische Universit\"at M\"unchen,
D-85748 Garching, Germany}\\

\normalsize

\end{center}
\vspace{0.3truecm}
\renewcommand{\baselinestretch}{1}
\begin{abstract}
We study the implications of a successful leptogenesis within the
framework of Minimal Lepton Flavour Violation combined with radiative
resonant leptogenesis and the PMNS matrix being the only source of CP
violation, which can be obtained provided flavour effects are taken
into account. We find that the right amount of the baryon asymmetry of
the universe can be 
generated in this framework with three quasi-degenerate heavy Majorana Neutrinos under the conditions of a normal hierarchy of
the light neutrino masses, a non-vanishing Majorana phase,
$\sin{(\theta_{13})}\gtrsim 0.13$ and $m_{\nu, {\rm lightest}}\lesssim
0.04$ eV. If this is fulfilled, we find strong correlations among
ratios of charged LFV processes.

\renewcommand{\baselinestretch}{1.2}
\end{abstract}

\thispagestyle{empty}

%
%
%
%
%
%

\thispagestyle{empty}

\setcounter{page}{1}
\pagenumbering{arabic}

\section{Preliminaries}
Leptogenesis \cite{Fukugita:1986hr} is an extremely successful mechanism to generate the observed
matter-antimatter asymmetry of the universe. While the common belief
in the past years 
was that distinguishing flavours in the Boltzmann equations is not
necessary, it turned out recently
\cite{Barbieri:1999ma, Endoh:2003mz, Pilaftsis:2005rv, Abada:2006fw, Abada:2006ea,Nardi:2006fx,
  Blanchet:2006be,Antusch:2006cw, Branco:2006hz,Antusch:2006gy} that flavour effects
can be relevant for the generation of the baryon asymmetry of the universe (BAU).
\par
For a long time, one thought that leptonic low-energy CP
violation does not automatically imply a non-vanishing
BAU through leptogenesis. This however does not
universally hold when ranges of the Majorana scale are considered in
which flavour effects play a role
\cite{Nardi:2006fx,Pascoli:2006ie,Pascoli:2006ci, Branco:2006ce,Abada:2006ea, Branco:2006hz}.
Leptogenesis without high-energy CP violation has been found to be successful in frameworks
with hierarchical heavy right-handed Majorana neutrinos \cite{Branco:2006ce,
  Pascoli:2006ie,Pascoli:2006ci}. But also in the case of
resonant leptogenesis \cite{Pilaftsis:2005rv, Pilaftsis:1997dr, Pilaftsis:2003gt, Blanchet:2006dq}, the BAU can be accommodated in the presence of exclusively
low-energy CP violation which has been shown
for a mass spectrum with two 
\cite{Pascoli:2006ci} and with three \cite{Branco:2006hz} quasi-degenerate heavy
right-handed Majorana neutrinos.
The analysis for three quasi-degenerate heavy right-handed Majorana
neutrinos which corresponds to the
Minimal Lepton Flavour Violation scenario and has been performed
in the un-flavoured regime by \cite{CIP06},
has been presented in \cite{Branco:2006hz} allowing for CP violation
at low and hight energies with the
BAU generated by radiative resonant leptogenesis (RRL) \cite{GonzalezFelipe:2003fi,Turzynski:2004xy,Branco:2005ye,Branco:2006hz}.\\
In the present paper, we concentrate on the limit of no high-energy
CP violation within this framework. We show that for a successful leptogenesis clear conditions
have to be fulfilled.\\
Applying these constraints we find strong correlations among low-energy lepton flavour
violating (LFV) decays, which are weak 
in the presence of high-energy CP violation \cite{Branco:2006hz}. Similar predictivity for ratios of LFV
decays has been observed in the single-flavour case in the limit of
large Majorana masses \cite{CIP06}.

\section{Minimal Lepton Flavour Violation}

The existence of neutrino masses implies that lepton flavour is not conserved.
However, from non-observation of LFV processes such as $\mu \to e
\gamma$ we know that those interactions have to be highly
suppressed. Extensions of the Standard Model (SM) that implement LFV
should keep such processes automatically small and allow for
new-physics particles with moderate masses.
In the quark sector, where the situation of flavour-changing
transitions is quite similar, these issues can nicely be accommodated
with the Minimal Flavour Violation (MFV) hypothesis \cite{MFV,D'Ambrosio:2002ex}. How this mechanism could be established in the lepton sector was proposed by \cite{CGIW}.
Analogously to the quark sector, Minimal Lepton Flavour Violation (MLFV)
can be formulated as an effective field theory in which the lepton Yukawa couplings  $Y_E$, $Y_{\nu}$ are the only sources of flavour violation. 
\be
\mathcal{L}_Y=-\bar e_R Y_E \phi^{\dagger} L_L   - 
\bar \nu_R Y_{\nu} \tilde\phi L_L  + h.c.
\ee
In order to additionally explain the smallness of neutrino masses with the
help of the see-saw mechanism, the MFV hypothesis in the lepton sector requires lepton number violation at some high scale, and three heavy right-handed Majorana neutrinos being introduced,
\be
\mathcal{L}_M=-\frac{1}{2} \bar \nu ^c_R M_R \nu_R + h.c. ~,
\ee
with the Majorana mass matrix having a trivial structure $M_R=M_\nu
\mathbbm{1}$. For a different definition of MLFV see \cite{Davidson:2006bd}. The branching ratios of LFV decays like $\mu \to e \gamma$ proceeding at a scale $\Lambda_{\rm LFV}$ in the framework of MLFV are given by \cite{CGIW}
\be\label{Brmutoe}
B(l_i \to l_j \gamma)=384 \pi^2 e^2 \frac{v^2}{\Lambda_{\rm LFV}^4}
|(Y_\nu^\dagger Y_\nu)_{ij}|^2 |C|^2.
\ee
Here $C$ summarizes the Wilson coefficients of the relevant operators
involved that can be calculated for a specified model. As the Wilson
coefficients are naturally of order one we will set those coefficients to
unity here. As $C$ is independent from the external lepton flavours,
this dependence is canceled when ratios of $l_i \to l_j \gamma$
decays are considered. In particular one has
 \be\label{RLFV}
R_{\rm LFV}=\frac{B(l_i \to l_j \gamma)}{B(l_m \to l_n \gamma)}=\frac{|(Y_\nu^\dagger Y_\nu)_{ij}|^2}{|(Y_\nu^\dagger Y_\nu)_{mn}|^2}.
\ee

\section{Radiative Resonant Leptogenesis}

Since radiative corrections spoil the degeneracy of
the Majorana masses \cite{CIP06, Branco:2006hz}, we combine the MLFV hypothesis with a choice of a scale at which the Majorana masses are exactly degenerate
as it was done in the analysis of \cite{Branco:2006hz}. A natural
selection for the degeneracy scale is the GUT scale $\Lambda_{\rm GUT}$,
\be
M_R(\Lambda_{\rm GUT})=M_\nu \mathbbm{1}.
\ee
The mass splittings at the Majorana scale that are necessary for leptogenesis are then induced {\it radiatively} which can be described by Renormalization Group Equations (RGE) and are approximately
\be \label{masssplittings}
\frac{\Delta M}{M} \sim Y_\nu Y_\nu^\dagger \ln{\left(\frac{M}{\Lambda_{\rm GUT}}\right)}.
\ee
For quasi-degenerate heavy right-handed Majorana neutrinos with mass splittings comparable to their decay widths, the CP asymmetries relevant for thermal leptogenesis are {\it resonantly} enhanced. The radiatively generated mass splittings (\ref{masssplittings}) automatically fulfill the condition of resonant leptogenesis \cite{Pilaftsis:1997dr, Pilaftsis:2003gt, Blanchet:2006dq}. For a more detailed description of {\it radiative resonant leptogenesis} and mass splittings due to the RGE of the SM and MSSM see \cite{Branco:2006hz}.

Thermal leptogenesis \cite{Fukugita:1986hr} requires CP and lepton
number violating out-of-equilibrium decays of heavy right-handed neutrinos. These decays produce a lepton asymmetry which is turned into a baryon asymmetry by sphaleron processes.
The CP asymmetry due to the Majorana neutrino $N_i$ and lepton flavour
$l$, defined as
\be        \label{eq:cpas}
\varepsilon_i^l=\frac{\Gamma(N_i\to L_l\,\phi)-\Gamma(N_i\to \bar L_l\, \bar \phi)}{\sum_l\left[\Gamma(N_i\to L_l\,\phi)+\Gamma(N_i\to \bar L_l \,\bar \phi)\right]}\,,
\ee
that enters the baryon asymmetry is given by
\be\label{CPA}
\varepsilon_i^l=\frac{1}{(Y_\nu Y_\nu^\dagger)_{ii}}\sum_{j} {\rm Im}((Y_\nu Y_\nu^\dagger)_{ij} (Y_\nu)_{il}(Y_\nu^\dagger)_{lj})\cdot g(M_i^2, M_j^2, \Gamma_j^2)
\ee
with the full expression of $g(M_i^2, M_j^2, \Gamma_j^2)$ given in
\cite{Pilaftsis:2003gt}. For a successful leptogenesis at the Majorana
scale in addition to $\Delta M\neq 0$, the Yukawa matrices have to contain
complex phases.

The neutrino Yukawa matrix $Y_\nu$ can be parametrized according to \cite{CAIB}\begin{equation}
\label{yukawaparamet}
 Y_{\nu}=\frac{i}{v}\,\sqrt{M_R}\, R \,\sqrt{m_{\nu}}
\,U_{\nu}^{\dagger}\, ,
\end{equation}
where $m_{\nu}=\text{diag}(m_{1},m_{2},m_{3})$ are the light-neutrino masses,
$M_R=\text{diag}(M_1,M_2,M_3)$ the masses of the heavy right-handed Majorana neutrinos, $U_{\nu}$ is the PMNS matrix and $R$ an orthogonal complex matrix
that encodes three physically relevant complex parameters \cite{Branco:2006hz}. Apart from radiative corrections, $R$ provides all high-energy CP
violation.\\
If $R$ is real, this corresponds to the limit of no high-energy CP
violation. In this case the only complex phases in $Y_\nu$ are those of the PMNS
matrix.

\section{Flavour Effects in Leptogenesis Scenarios}

There have been several investigations recently of the importance of
flavour effects
\cite{Barbieri:1999ma, Pilaftsis:2005rv,Abada:2006fw,Nardi:2006fx,
  Blanchet:2006be, Antusch:2006cw, Branco:2006hz}. Below some temperature, the
interactions associated with the $\mu$ and $\tau$ charged lepton
Yukawa couplings are much faster than the expansion of the universe
and so are in equilibrium. This is the case for $T \sim M_\nu \lesssim 10^9-10^{12}$
GeV. The interactions of the $\tau$ Yukawa coupling are in equilibrium
below about $10^{12}$ GeV, where two flavours then can be distinguished,
followed by the ones of the $\mu$ Yukawa coupling below
approximatively $10^9$ GeV, where three distinguishable flavours exist.
Also processes that wash out the lepton asymmetries are flavour
dependent. In these regimes below $10^{9} -10^{12}$ GeV, flavour
specific solutions to the Boltzmann equations are required.
\par
An order-of-magnitude estimate of solutions for the BAU $\eta_B$ of the Boltzmann equations in the strong washout regime in the single-flavour treatment is given by~\cite{Pilaftsis:2005rv} 
\be\label{etaB-estimate_nf}
\eta_B\simeq -10^{-2} \,\sum_{i=1}^3\,e^{-(M_i-M_1)/M_1}\,\frac{1}{K} \,\sum_{l=e,\mu,\tau}
\varepsilon_i^l,
\ee
 and an estimate including flavour effects by \cite{Pilaftsis:2005rv} 
\be\label{etaB-estimate}
\eta_B\simeq -10^{-2} \,\sum_{i=1}^3\sum_{l=e,\mu,\tau}\,e^{-(M_i-M_1)/M_1}\,
\varepsilon_i^l\,\frac{K_i^l}{K^l K_i},
\ee
with 
\be
K_i^l=\frac{\Gamma(N_i\to L_l \phi)+\Gamma(N_i \to \bar L_l\bar\phi)}{H(T=M_i)}
\ee
\be
K = \sum_i K_i, \qquad K_i=\sum_{l=e,\mu,\tau}K_i^l,\qquad K^l=\sum_{i=1}^3 K_i^l, \qquad
H(T=M_i)\simeq 17 \frac{M_i^2}{M_{\rm Pl}} .
\ee
Here $M_{\rm Pl}=1.22\times 10^{19}$ GeV and $K_i^l$ is the washout factor due to the inverse decay of the Majorana
neutrino $N_i$ and the lepton flavour $l$. In the strong washout
regime the following conditions for
the single-flavour (\ref{1flav}) and the three-flavour (\ref{3flav}) case have to be fulfilled:
\be\label{1flav}
K = K_1 + K_2 + K_3  \gtrsim 50,
\ee
\be\label{3flav}
K_i^l\gtrsim 1.
\ee
Similar estimates can be
found in \cite{Blanchet:2006be,Blanchet:2006dq}.

In the past, the common belief was that for a successful leptogenesis
high-energy CP violation is required since the complex phases from the
PMNS matrix do not offer a sufficient amount of CP violation. This
statement is not true when flavour effects are considered. A
flavour-specific treatment of the washout and the CP asymmetries can
accommodate the BAU of the right order of magnitude
\cite{Pascoli:2006ie,Pascoli:2006ci, Branco:2006ce, Branco:2006hz}.
When flavours have to be treated specifically, the un-summed terms in
the CP asymmetry (\ref{CPA}), $(Y_\nu)_{il}(Y_\nu^\dagger)_{lj}$, can
become important. Summing over the flavours as in the estimate
(\ref{etaB-estimate_nf}), it follows from (\ref{yukawaparamet}) that the
PMNS matrix cancels due to its unitarity (apart from radiative
corrections between the GUT and the Majorana scale) while this is not the case in
(\ref{etaB-estimate}), where the CP asymmetry of a single flavour is
weighted separately due to the $K$ factors. This gives a notion how
the PMNS phases come
into play when flavours are treated specifically.

\section{Leptogenesis as a Constraint of LFV Processes}

Since the size of $B(l_i \to l_j \gamma)$ is governed by the size of
the ratio $ M_\nu^2/\Lambda_{\rm LFV}^4$ in the
context of MLFV, there is in principle a rich spectrum of
possibilities for the size of such effects. The question is whether a
successful generation of the BAU by radiative resonant leptogenesis
provides a strong correlation with LFV processes. 
The observed value of the BAU is
given by \cite{Spergel:2006hy}
\be
\eta_B=\frac{n_B}{n_\gamma}=(6.10 \pm 0.21) \cdot 10^{-10}.
\ee
Even if the baryon asymmetry is just a single number, it provides an
important relation between the early-universe cosmology and the SM
of particle physics and its extensions. 
Successful leptogenesis could constrain the large amount of
possibilities for $l_i \to l_j \gamma$ or ratios of such
decays. However, the more parameters are contained in a given theory, the
less restrictive is the constraint that stems from the baryon asymmetry.

\subsection{Low- and high-energy CP Violation}

Recently, two analyses were performed that implemented leptogenesis in the
framework of MLFV and included CP violation at low and high energies.
In \cite{CIP06} the BAU was generated in a framework similar to
RRL. In this analysis, flavour effects were not considered. In this case, the
BAU provides a lower mass bound on the Majorana scale of $\sim
10^{12}$ GeV. In this framework a relation among LFV processes could be obtained:
\be\label{R}
B(\mu \to e \gamma)<B(\tau \to \mu \gamma).
\ee
In \cite{Branco:2006hz}  with the inclusion of flavour effects in the
relevant regimes, the
BAU of the right order of magnitude could be accommodated in the
framework of RRL with RGE of the SM and MSSM basically independent from the
Majorana scale. Due to this fact, it turned out that there are only
very weak correlations among leptogenesis and charged LFV processes and the relation
(\ref{R}) could not be confirmed. In contrast to MFV in the quark sector,
predictivity seems to be almost lost in MFV in the lepton sector when
high-energy CP violation is present and flavour effects are considered. \\

\subsection{Low-Energy CP Violation alone}

\begin{figure}[h]
\vspace{0.01in}
\centerline{
\epsfysize=2in
\epsffile{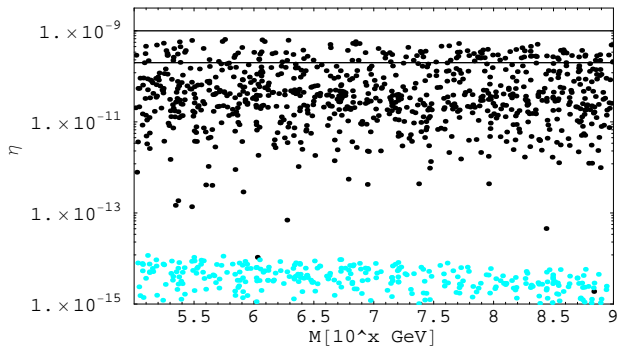}}
\vspace{-0.18in}
\caption{The BAU $\eta_B$ versus the Majorana scale up to $10^9$ GeV.
  The black points correspond to the three-flavour estimate, the light-blue points to the single-flavour
  solution. The two black lines mark where $\eta_B$ is of the right order
  of magnitude.}\label{fig:etaM}
\end{figure}

In this section we consider the framework of MLFV and RRL in
the three-flavour regime ($M_\nu<10^9$ GeV) 
with RGE for the SM and investigate conditions for a successful leptogenesis without high-energy CP violation.\\
For the PMNS matrix that describes leptonic low-energy mixing and CP
violation, we use
the convention
\be\label{standard}
U_\nu=
\left( 
\begin{array}{ccc}
c_{12}c_{13} &s_{12}c_{13} &s_{13}e^{-i\delta} \\
-s_{12}c_{23}-c_{12}s_{23}s_{13}e^{i\delta} &c_{12}c_{23}-s_{12}s_{23}s_{13}e^{i\delta}  & s_{23}c_{13} \\
s_{12}s_{23}-c_{12}c_{23}s_{13}e^{i\delta} & -s_{23}c_{12}-s_{12}c_{23}s_{13}e^{i\delta}  & c_{23}c_{13} 
\end{array}
\right)\cdot V
\ee
where $c_{ij}=\cos(\theta_{ij})$, $V={\rm diag}(e^{i\alpha}, e^{i\beta},1)$,  $\alpha$ and
$\beta$ denote the Majorana phases and $\delta$ denotes the Dirac
phase. $0 \leq \alpha, \beta \leq \pi$  and $0 \leq \delta \leq
2 \pi$ are the physically
relevant ranges for the PMNS phases. PMNS matrix elements enter the CP
asymmetries according to (\ref{CPA}) and
(\ref{yukawaparamet}). Further we consider the following ranges:
$10^5$ GeV $<M_\nu<10^{9}$ GeV in which the three-flavour estimate of
$\eta_B$ (\ref{etaB-estimate}) can be applied, $0\leq
\sin{(\theta_{13})}\leq 0.2$ and $0\leq
m_{\nu, {\rm lightest}}\leq 0.2$ eV. We use $c_{23}=s_{23}=1/\sqrt{2}$
and $\theta_{12}=33^\circ$ and for the light neutrinos we have the low energy values 
\be\label{In1}
\Delta m^2_{\rm sol}=m_{ 2}^2-m^2_{ 1}=8.0\cdot 10^{-5}~\ev^2
\ee
\be\label{In2}
\Delta m^2_{\rm atm}=| m_{3}^2-m^2_{ 2}|=2.5\cdot 10^{-3}~\ev^2
\ee
with $m_{\nu, {\rm lightest}}=m_{1}(m_{3})$ for normal (inverted) hierarchy,
respectively.
\par
In the limit of vanishing high-energy CP violation, the matrix $R$ of (\ref{yukawaparamet})
has to be real and we set it equal to unity at the GUT scale since at
this scale the heavy right-handed Majorana neutrinos are exactly degenerate.
\be
R(\Lambda_{\rm GUT})=\mathbbm{1}
\ee
Then the only complex phases that enter the neutrino Yukawa matrix
$Y_\nu$ and so the BAU apart from radiative corrections are the one of the
PMNS matrix. 
\begin{figure}[h]
\vspace{0.01in}
\centerline{
{\epsfysize=2in
\epsffile{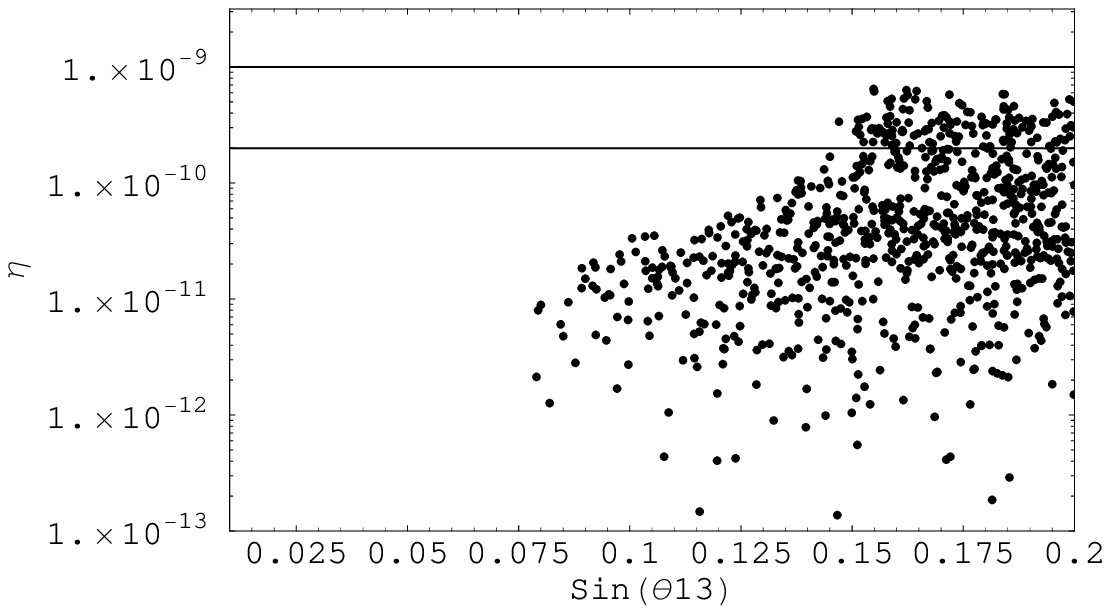}}{\epsfysize=2in\epsffile{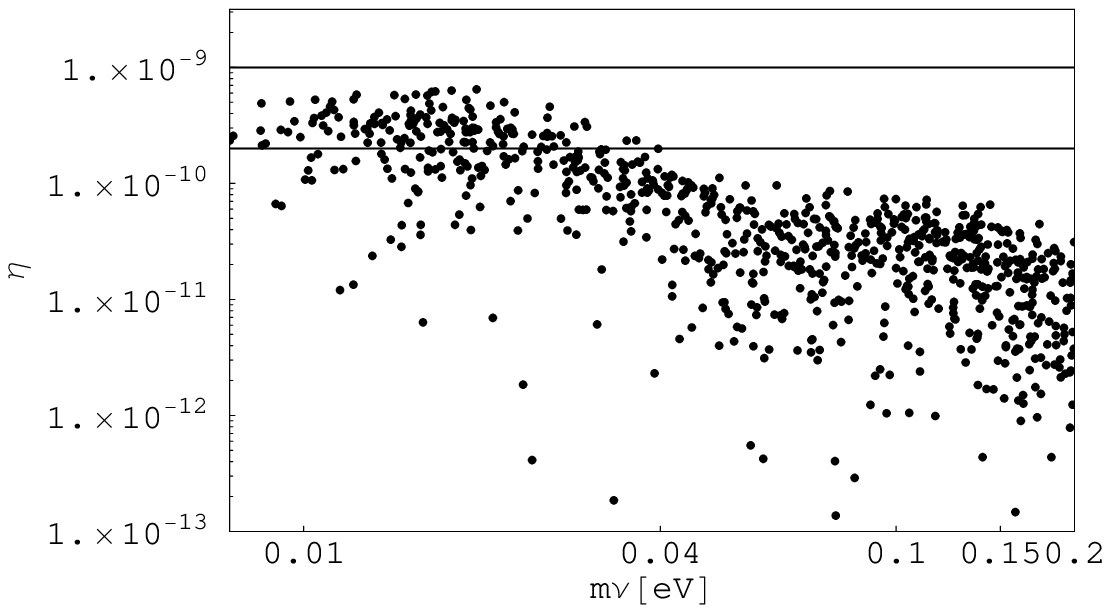}}}
\vspace{-0.18in}
\caption{The BAU $\eta_B$ versus $\sin{(\theta_{13})}$ and the lightest
  neutrino mass $m_\nu$. In our scenario with exclusively low-energy
  CP violation, successful leptogenesis
  requires $0.13\lesssim \sin{(\theta_{13})}$ and $m_\nu
  \lesssim 0.04$ eV.}\label{fig:t13m1}
\end{figure}
It has been found that it is possible to obtain the baryon asymmetry generated
by RRL with only low-energy CP
violation of the right order of magnitude in the regime where the
flavour-dependent estimate (\ref{etaB-estimate}) is valid
\cite{Branco:2006hz}. This can also be
seen in Figure \ref{fig:etaM}, showing that the BAU
can be accommodated properly independent from the Majorana scale with
the estimate (\ref{etaB-estimate}), whereas with a single-flavour treatment
in this regime, the right size of the BAU cannot be obtained
(light-blue points).\\
In the scenario presented here with exclusively low-energy
  CP violation, we find that successful leptogenesis implies the lower
  bound
\be
\sin{(\theta_{13})}\gtrsim 0.13.
\ee
For the lightest neutrino
  mass we obtain the upper bound
\be
 m_\nu
  \lesssim 0.04\; {\rm eV}
\ee
(see Figure \ref{fig:t13m1}). These bounds can be tested experimentally
providing information on the viability of this framework.

If not stated differently, all plots presented here correspond to {\it normal
  hierarchy} in the light neutrino masses. 
\par
\begin{figure}[h]
\vspace{0.01in}
\centerline{
{\epsfysize=2in
\epsffile{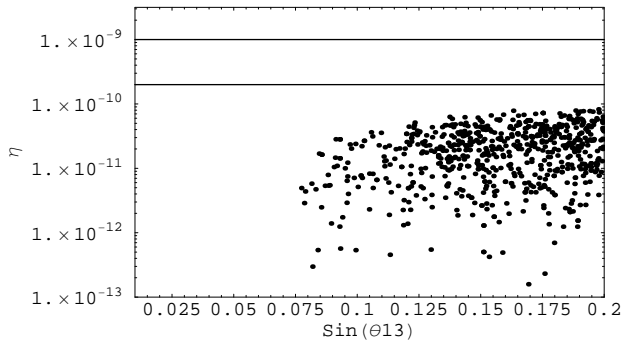}}}
\vspace{-0.18in}
\caption{The BAU $\eta_B$ versus $\sin{(\theta_{13})}$ for {\it inverted
  hierarchy} of the light neutrino masses with exclusively low-energy CP
violation. It is not possible to obtain the BAU of the right size with
{\it inverted hierarchy}.}\label{fig:inverted}
\end{figure}

For a scenario with two quasi-degenerate heavy right-handed neutrinos, it
has been found \cite{Pascoli:2006ci} that is is possible to generate
the BAU of the right order of magnitude for all hierarchies of the
light neutrino masses. \\
In this setup with three quasi-degenerate heavy right-handed neutrinos, we find that for {\it inverted hierarchy},
 it is not possible to generate the BAU of the right
order of magnitude with low-energy CP violation alone and RGE for the SM. This is shown
in Figure \ref{fig:inverted}.

\subsubsection{CP Violation governed by a single PMNS phase}
\begin{figure}[h]
\vspace{0.01in}
\centerline{
{\epsfysize=2.in
\epsffile{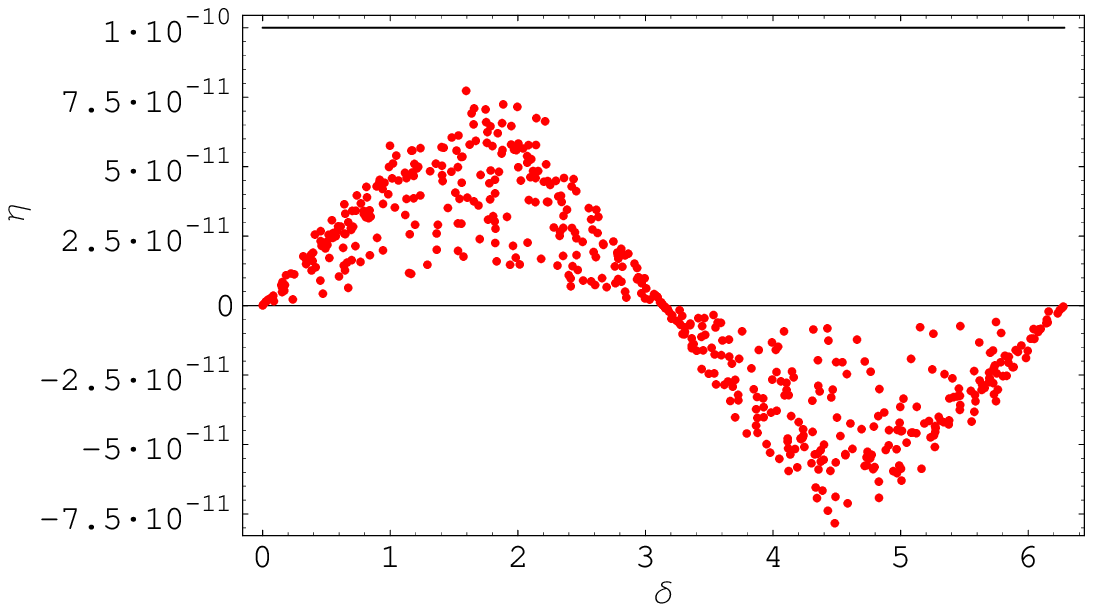}}\hspace{0.15cm}{\epsfysize=2.in\epsffile{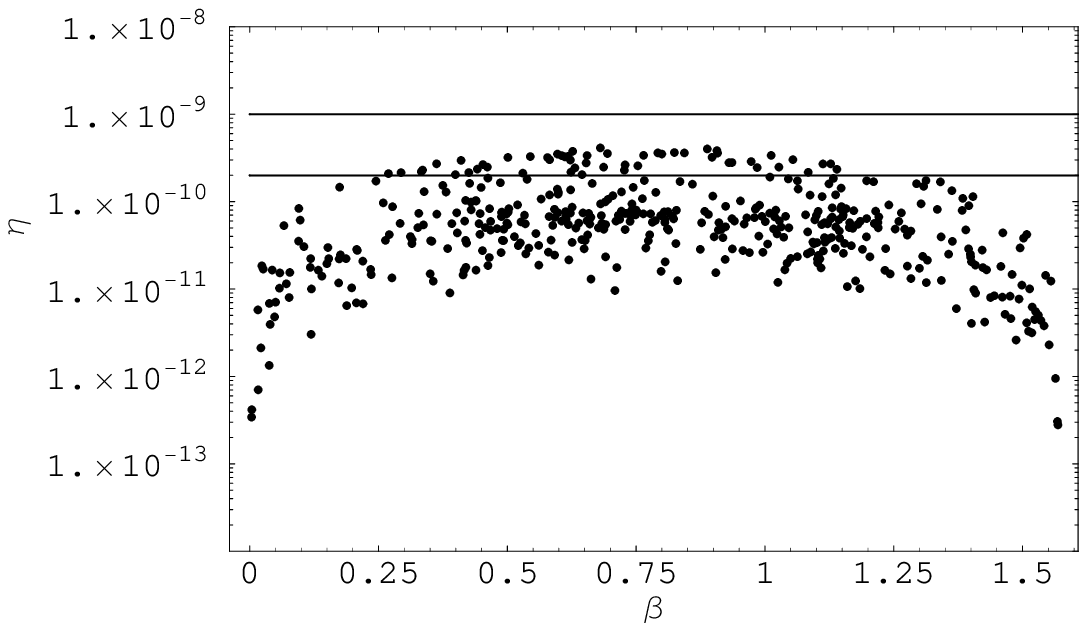}}}
\vspace{-0.18in}
\caption{Left Plot: The baryon asymmetry $\eta_B$ (left) plotted over the
  Dirac phase $\delta$ for  $\delta$ being the only source of  CP
  violation which is not sufficient to obtain the right
order of magnitude of the BAU indicated by the black line in the left plot. The range $\pi \leq \delta \leq 2 \pi$
corresponds to negative values of $\eta_B$.
Right Plot: The baryon asymmetry $\eta_B$ with the Majorana phase $\beta$
  being the only source of CP violation.}\label{fig:JCP}\label{fig:beta}
\end{figure}

Now we want to investigate whether single CP violating phases could be sufficient
to generate the BAU.
If the Dirac phase $\delta$ is the only complex phase involved (see
left plot of Figure \ref{fig:JCP}), we find
that it is not possible to fulfill the
leptogenesis constraint which can be explained by the suppression by $\sin (\theta_{13})$ of the corresponding PMNS entries.
This implies that the observation of CP violating neutrino
oscillations alone is not sufficient to ensure successful leptogenesis
in this framework.\\
However it is possible to successfully generate the BAU with a single
Majorana phase $\alpha$ or $\beta$. This is depicted in the right plot
of Figure
\ref{fig:beta} where $\beta \neq 0$, $\alpha=\delta=0$ and
$R(\Lambda_{\rm GUT})=\mathbbm{1}$.
This corresponds to the results of \cite{Pascoli:2006ci} where in the resonant case
a strong sensitivity on the Majorana phases has been observed.\\
Therefore one can say that experimental observation or non-observation
of Majorana phases will decide whether the setup presented could be
realized in nature.

\subsubsection{LFV Processes}

From equation (\ref{Brmutoe}) one can read that $B(\mu \to e \gamma)$
rises with increasing Majorana scale for a fixed scale of lepton
number violation.
In a scenario with low-energy CP
violation the Majorana scale is bounded by up to which scale the
flavour-dependent analysis is valid. In this setup we chose
 $10^9$ GeV for to be conservative. Analyses that include the two-flavoured regime
$10^9 \; {\rm GeV}<M_\nu<10^{12}$ GeV have to decide whether successful leptogenesis
without high-energy CP violation for larger Majorana scales is possible.\\
\begin{figure}[h]
\vspace{0.01in}
\centerline{
{\epsfysize=2in
\epsffile{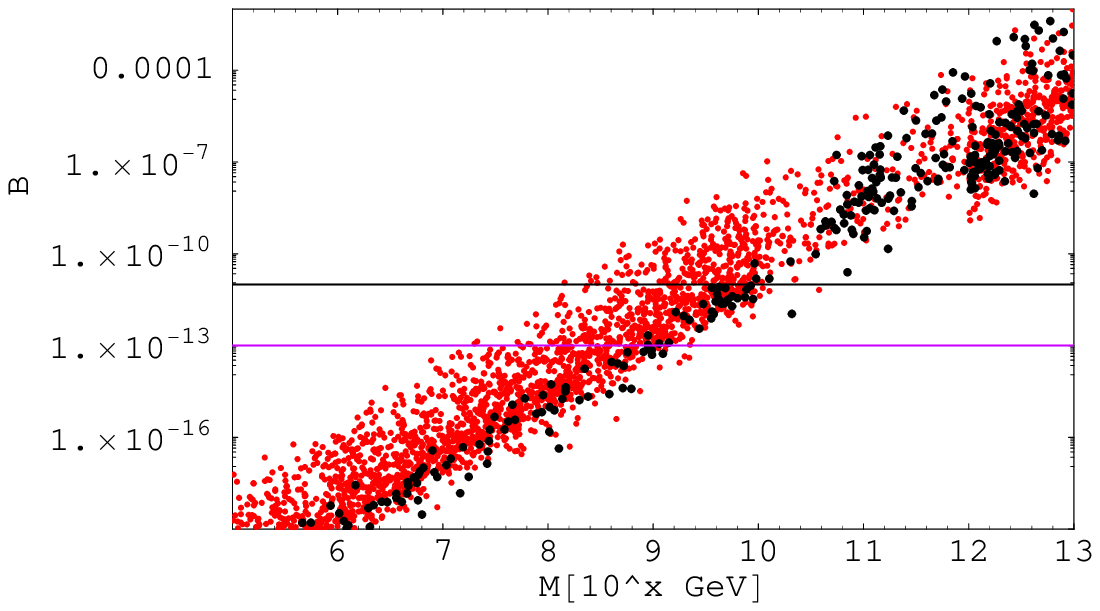}}{\epsfysize=2in\epsffile{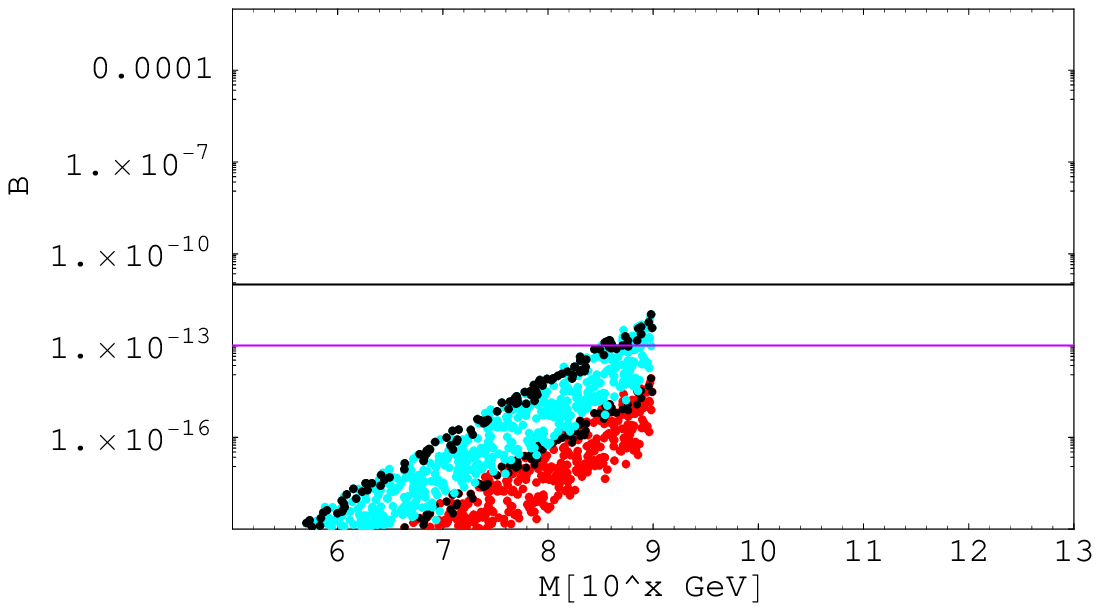}}}
\vspace{-0.18in}
\caption{$B(\mu \to e \gamma)$ versus
  $M_\nu$ for the
  general analysis \cite{Branco:2006hz} including high-energy CP
  violation and $\Lambda_{\rm LFV}=1$ TeV (left) and without
  high-energy CP violation (right) where $\Lambda_{\rm LFV}=1$ TeV
  (red) and  $\Lambda_{\rm LFV}=300$ GeV (light-blue). The black points
  indicate where the leptogenesis constraint is fulfilled. The black
  line corresponds to the present bound on $B(\mu \to e
  \gamma)<10^{-11}$ and the lower pink line to the sensitivity of the upcoming
  PSI experiment $\sim 10^{-13}$.  }\label{fig:Bmue}
\end{figure}
\begin{figure}[h]
\vspace{0.01in}
\centerline{
{\epsfysize=2in
\epsffile{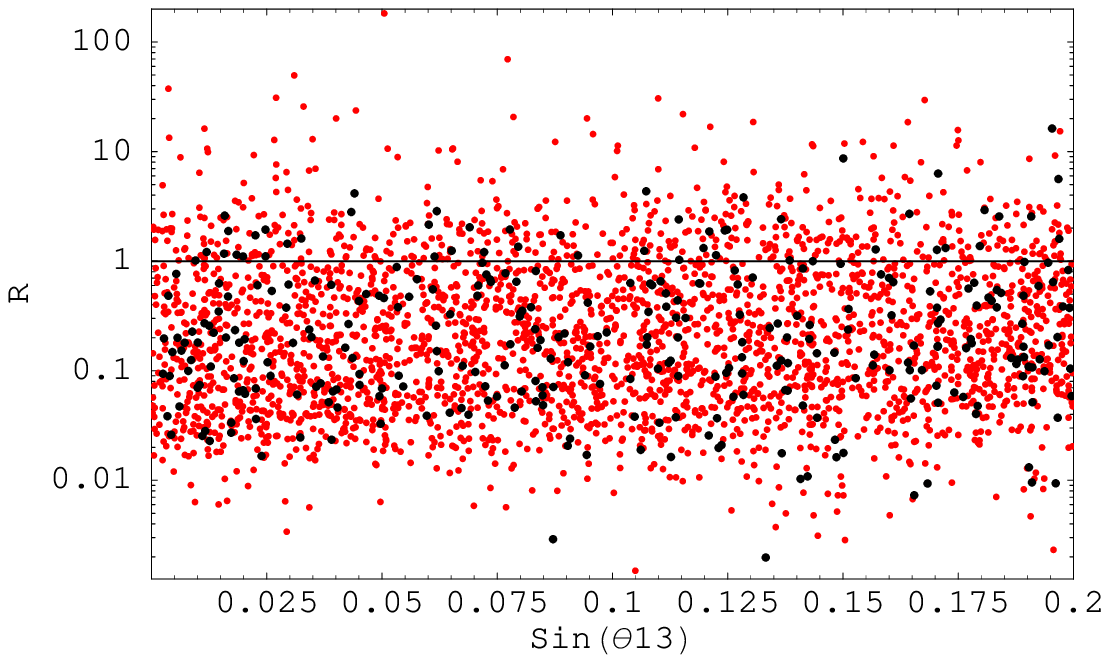}}{\epsfysize=2in\epsffile{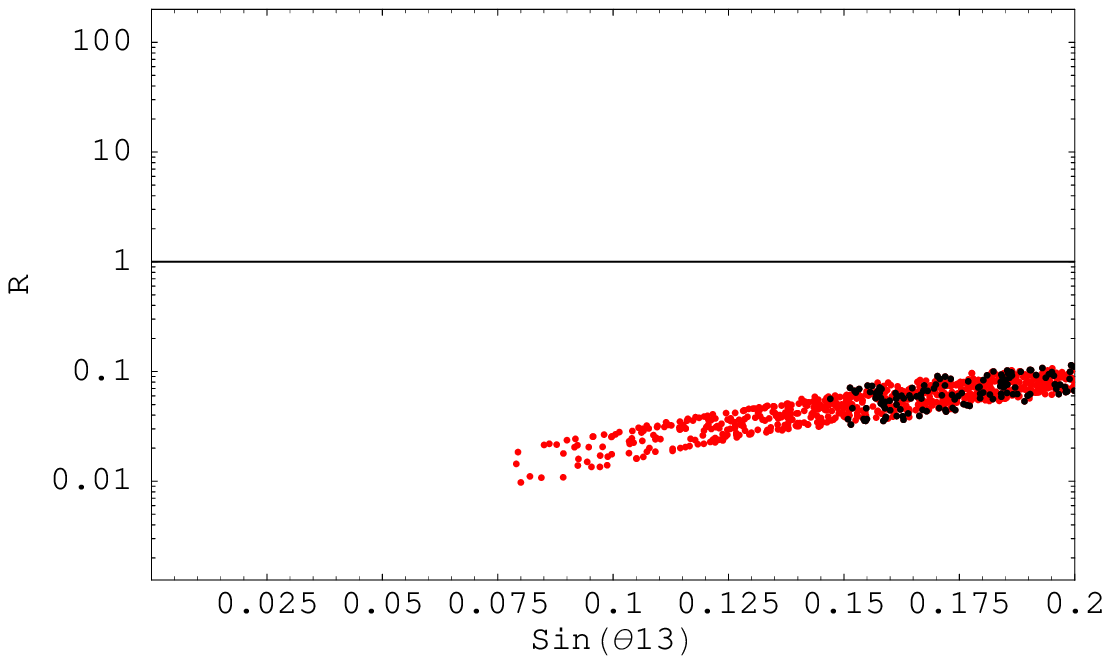}}}
\vspace{-0.18in}
\caption{$R_{\rm LFV}=B(\mu \to e \gamma)/B(\tau \to \mu \gamma)$ versus
  $\sin{(\theta_{13})}$ for the
  general analysis \cite{Branco:2006hz} including high-energy CP
  violation (left) and without high-energy CP violation (right) where
  the relation (\ref{R}) is satisfied. The black points fulfill the
  leptogenesis constraint.}\label{fig:R}
\end{figure}
If this is not the case and the Majorana scale is bounded to be below
$10^9$ GeV as in the scenario considered,
$\Lambda_{\rm LFV}$ could be as low as 300-500 GeV to obtain $B(\mu
\to e \gamma)$ in the
reach of the PSI experiment (see Figure \ref{fig:Bmue}).
\par
The very important point of this analysis is that with exclusively
low-energy CP violation the relation $B(\mu \to e \gamma)<B(\tau \to
\mu \gamma)$ is valid which was not
true in the case of CP violation at low and high energies
\cite{Branco:2006hz} (see Figure \ref{fig:R}). Furthermore we find  $B(\tau \to e \gamma)<B(\tau \to
\mu \gamma)$. Figure \ref{fig:R} nicely
depicts how dramatic the situation changes when high-energy CP
violation is excluded. The rich amount of possibilities for the
ratio $R_{\rm LFV}=B(\mu \to e \gamma)/B(\tau \to
\mu \gamma)$ that is present in the high-energy CP violation case even
if it solely depends on the relevant Yukawa couplings (\ref{RLFV}), can then
significantly be constrained.  
We find MLFV with low-energy
CP violation predicts a strong correlation among these LFV decays that
can be tested in the future.

\section{Conclusions}

Recent studies showed the relevance of the inclusion of flavour effects in the
Boltzmann equations for the generation of the baryon
asymmetry of the universe and the existence
in the flavoured-regime CP violation at high energies is no longer a
necessary requirement for successful leptogenesis.
\par
When leptogenesis is implemented in the framework of Minimal Lepton
Flavour Violation by radiative resonant leptogenesis including
high-energy CP violation \cite{Branco:2006hz}, a rich spectrum
of possibilities for charged LFV processes and ratios of such
processes is present matching with the leptogenesis constraint
implying that 
correlations among leptogenesis and LFV are weak.
\par
In this paper we analysed the framework of Minimal Lepton Flavour Violation in the limit of no CP violation at high
energies, with the PMNS phases being the only complex ingredient apart
from radiative corrections. 
We find that with radiative resonant leptogenesis with RGE of the SM, one
can successfully generate the BAU in the three-flavour regime provided that:
\begin{itemize}
\item there is a non-vanishing Majorana phase, 
\item the light neutrino masses have normal hierarchy,
\item the lightest neutrino mass and the PMNS angle $\theta_{13}$ fulfill
   $m_{\nu}\lesssim 0.04$ eV and
 $0.13 \lesssim
\sin{(\theta_{13})}$ .
\end{itemize}
When these constraints are fulfilled, we find
strong correlations among LFV processes and that the rich spectrum of
possibilities that is present when high-energy CP violation is
included \cite{Branco:2006hz} can significantly
be reduced.
In this case MLFV 
turns out to be much more
predictive.

\noindent
{\bf Acknowledgments}\\
\noindent
 We would like to particularly thank Sebastian J\"ager and Andreas
 Weiler for discussions that lead to the idea of the paper,
for their contributions to the framework considered and to the
computer programs used.
Further we thank Steve Blanchet, Gustavo C. Branco, Andrzej J. Buras, Pasquale
Di Bari, Ulrich Haisch, Gino Isidori and Stefan Recksiegel for
discussion.  We would like to thank Andrzej Buras for comments on the manuscript.
This work has been supported by
Bundesministerium f\"ur
Bildung und Forschung under the contract 05HT6WOA.

\renewcommand{\baselinestretch}{0.95}

\end{document}